\documentclass[11pt, a4paper]{article}
\usepackage[top=30truemm,bottom=30truemm,left=20truemm,right=20truemm]{geometry}

\usepackage{amsmath}
\usepackage{amssymb}
\makeatletter
    
    \@addtoreset{equation}{section}
\makeatother

\title{\vbox{%
\baselineskip 14pt
\hfill \hbox{\normalsize KUNS-2443}\\
} \vskip 1.7cm
\Large \bf Quantum 3D Tensionless String in Light-cone Gauge 
\vskip 0.5cm
}
\author{ \\
 \\
Kenta Murase$^1$\thanks{E-mail: kmurase@gauge.scphys.kyoto-u.ac.jp or kenta1murase2@gmail.com}\\
 \\
{\sl\fontsize{10pt}{0pt}\selectfont $^1$Department of Physics, Kyoto University, Kitashirakawa, Kyoto 606-8502, Japan }\\ }
\date{}

\begin{document}
\maketitle

\begin{abstract}
We discuss the quantization of a tensionless closed string in light-cone gauge. 
It is known that by using a Hamiltonian BRST scheme a tensionless string has no Lorentz anomaly in any space-time dimensions and no anomaly for the space-time conformal symmetry in two dimensions.  
In this paper, we show that a 3d tensionless closed string in light-cone gauge also has no anomaly of space-time conformal symmetry. 
We also study the spectrum of a 3d tensionless closed string.
\end{abstract}

\newpage

\tableofcontents

\section{Introduction}
It is well known that the critical dimension of bosonic (or supersymmetric) string theory is 26 (or 10) and we can check this fact by many methods, including Light-cone quantization and BRST quantization \cite{bib:001, bib:002}. 
Recently, Luca Mezincescu and Paul K. Townsend showed that by using light-cone gauge a consistent critical string theory can be constructed also in three dimensions, since there is no Lorentz anomaly in three dimensions.
In fact, in three dimensions the dangerous commutator which breaks Lorentz symmetry, $[\mathcal{J}^{-I}, \mathcal{J}^{-J}] \ (I,J=2, \cdots ,D-1)$, vanishes trivially because there is only one transverse direction \cite{bib:003, bib:004, bib:005, bib:021}.
\begin{eqnarray}
[\mathcal{J}^{-I}, \mathcal{J}^{-I}] \equiv 0 ~~~ \mbox{in 3 dim.}
\label{eq:1-1}
\end{eqnarray}
Moreover they found that the spectrum of a 3d string in light-cone gauge includes "anyons", which have non half-integer spins. 

The difference between the critical dimension in the light-cone gauge quantization and that in others might mean the fault of the light-cone quantization or the incompleteness of other quantizations including BRST method.
Mezincescu and Townsend suggest that such difference may be caused by the existence of anyon, although it is still not clear whether this is in the case and how the difference arises.
Beside finding a reason for the difference, it is also important to find other examples which give a different result in the light-cone method and others and to invent other quantization schemes which reproduce results obtained by the light-cone method, especially by the covariant one.

In this paper, we investigate 3d tensionless bosonic closed string. 
It has been known by using BRST method that tensionless p-branes have no Lorentz anomaly in any dimensions and that conformal tensionless p-branes have the critical dimension for the space-time conformal symmetry, $D=2$ \cite{bib:013, bib:014, bib:015, bib:006}. 
And the mass spectrum also has been investigated in \cite{bib:016, bib:020}.

On the other hand, Isberg, Lindstrom, Sundborg and Theodoridis show that there are some anomalous commutators in the space-time conformal group of light-cone gauge \cite{bib:007, bib:017}.
Specifically, $[\mathcal{K}^{I}, \mathcal{J}^{-J}]$, $[\mathcal{K}^{+}, \mathcal{K}^{-}]$, $[\mathcal{K}^{I}, \mathcal{K}^{-}]$ and $[\mathcal{K}^{-}, \mathcal{J}^{-I}]$ have anomalous terms. 
Especially Isberg et al. emphasize in \cite{bib:007, bib:017} that the commutator $[\mathcal{K}^{I}, \mathcal{J}^{-J}]$ has anomalous terms in traceless part of $I$ and $J$ and gives the definition of $\mathcal{K}^{-}$ by trace part:
\begin{eqnarray}
[\mathcal{K}^{I}, \mathcal{J}^{-J}] = -i\delta ^{IJ} \mathcal{K}^{-} +L^{IJ}_{t.l.},
\label{eq:1-2}
\end{eqnarray}
where $L^{IJ}_{t.l.}$ is traceless and proportional to the generator of $\sigma $-dependent special linear group.
In $D\geq 4$, we will find that through Jacobi identities all other anomalous commutators are related to this commutator or the definition of $\mathcal{K}^{-}$.
For example, if we choose $J$ different from $I$ and use the fact that there is no Lorentz anomaly for a tensionless string\footnote{$[\mathcal{J}^{-I}, \mathcal{J}^{-J}]\propto $($\sigma $-translation gauge-fixing constraint). This constraint corresponds to the level-matching condition for tensile string and commutes to all generators.}, we obtain 
{\setlength\arraycolsep{1pt}\begin{eqnarray}
[\mathcal{K}^{-}, \mathcal{J}^{-I}] &=& i[ [\mathcal{K}^{J}, \mathcal{J}^{-J}]-L^{JJ}_{t.l.}, \mathcal{J}^{-I}] \nonumber \\
&=& -i[ [\mathcal{J}^{-I}, \mathcal{K}^{J}], \mathcal{J}^{-J}] -i[ [\mathcal{J}^{-J}, \mathcal{J}^{-I}], \mathcal{K}^{J}] -i[L^{JJ}_{t.l.}, \mathcal{J}^{-I}] \nonumber \\
&=& i[L^{JI}_{t.l.}, \mathcal{J}^{-J}]-i[L^{JJ}_{t.l.}, \mathcal{J}^{-I}].
\label{eq:1-3}
\end{eqnarray}}
Similarly we will find that all anomalies derive from $L^{IJ}_{t.l.}$.

In three dimensions, we can readily find that no anomalous terms appear in $[\mathcal{J}^{-I}, \mathcal{K}^{J}]$, that is, $L^{IJ}_{t.l.}=0$. 
Using Jacobi identities and $[\mathcal{K}^{+}, \mathcal{K}^{I}]=0$ which is checked easily, we obtain $[\mathcal{K}^{+}, \mathcal{K}^{-}]=0$. 
Because we can't set $I\not= J$ in three dimensions, however, $[\mathcal{K}^{-}, \mathcal{J}^{-I}]$ and $[\mathcal{K}^{I}, \mathcal{K}^{-}]$ are not related to $L^{IJ}_{t.l.}$. 
We just get the relation between $[\mathcal{K}^{-}, \mathcal{J}^{-I}]$ and $[\mathcal{K}^{I}, \mathcal{K}^{-}]$\footnote{$[\mathcal{K}^{-}, \mathcal{J}^{-I}]=-i[\mathcal{P}^{-}, [\mathcal{K}^{-}, \mathcal{K}^{I}]]$ or $[\mathcal{K}^{I}, \mathcal{K}^{-}]=-i[[\mathcal{J}^{-I}, \mathcal{K}^{-}], \mathcal{K}^{+}]$}. 
Therefore it is nontrivial whether $[\mathcal{K}^{-}, \mathcal{J}^{-I}]$ is anomalous and we need to investigate it in order to conclude the absence of space-time conformal anomaly.
Fortunately, because the vanishing of $[\mathcal{K}^{-}, \mathcal{J}^{-I}]$ is equivalent to that of $[\mathcal{K}^{I}, \mathcal{K}^{-}]$, we just have to check $[\mathcal{K}^{-}, \mathcal{J}^{-I}]=0$.
This check is the first content of this paper.

In this paper, we write operators by the Fourier expansion and choose the "reference ordering"\footnote{In the reference ordering all P-modes and $p_{-}$ are to the right of all X-modes and $x^{-}$\cite{bib:007, bib:017}.}.
Then we investigate commutators of the space-time conformal group for Hermitian generators.
In the calculation, we need to introduce some regularization because we find many troublesome divergences. 
In this paper, we regularize by removing higher modes of operators than a given cutoff and find $[\mathcal{K}^{-}, \mathcal{J}^{-I}]=0$.

As the second content of this paper, we also investigate the spectrum of a 3d tensionless bosonic closed string.
We get massive states and massless states there.
Because massless states are conformal invariant, we discuss it in more detail.

The content of the paper is as follows: In Section 2, we represent a 3D tensionless closed bosonic string in light-cone gauge.
In section 3, we quantize a 3D tensionless closed bosonic string in light-cone gauge and find that this string theory has the space-time conformal symmetry. In Section 4, we discuss the spectrum of a 3d tensionless string. In Section 5, we end the paper with the conclusion and outlooks. The definition of light-cone coordinate and the algebra of conformal group are collected in appendices.

\section{3D tensionless string in light-cone gauge}
In this section we consider in three dimensions the light-cone quantization of a string without tension, namely a tensionless string. 
We follow the method of \cite{bib:003} to quantize a tensionless string.

First we consider a string with tension, namely a tensile string.
A 3D bosonic closed string with tension $T$ is described by Nambu-Goto action:
\begin{eqnarray}
S[\mathbf{X}]=-T\int d\tau \oint \frac{d\sigma }{2\pi } \sqrt{\left( \left( \dot{\mathbf{X}} \cdot \mathbf{X}' \right) ^2 -\dot{\mathbf{X}} ^2 (\mathbf{X}')^2 \right) } ,
\label{eq:2-1}
\end{eqnarray}
where ${\mathbf{X}^{\mu }(\tau ,\sigma );\mu =0,1,2}$ represents an embedding of the world sheet ${(\tau ,\sigma )}$ to 3D Minkowski space with matric $\eta =\mbox{diag}(-1,1,1)$. An overdot indicates a derivative with respect to time parameter $\tau $ and a prime indicates a derivative with respect to string coordinate $\sigma $.
The centerdot or superscript "2" indicate the contraction.
Moreover we assume that the functions are periodic, $X^{\mu }(\tau , \sigma ) = X^{\mu }(\tau , \sigma +2\pi )$.

By using the conjugate momentums $\mathbf{P}_{\mu }$, and the auxiliary fields $V$ and $U$, we can rewrite the action (\ref{eq:2-1}) to the form
\begin{eqnarray}
S[\mathbf{X},\mathbf{P};V,U] = \int d\tau \oint \frac{d\sigma }{2\pi } \left\{ \dot{\mathbf{X}} ^{\mu } \mathbf{P}_{\mu } -\frac{1}{2} V\left[ \mathbf{P}^2 +(T\mathbf{X}') ^2 \right] -U \mathbf{X}'^{\mu }\mathbf{P}_{\mu } \right\} ,
\label{eq:2-2}
\end{eqnarray}
where $V$ and $U$ are the Lagrange multipliers for the Hamiltonian and $S^{1}$-diffeomorphism constraints, respectively.
When one eliminates $\mathbf{P}_{\mu }$, followed by the elimination of $V$ and $U$ in order, the Nambu-Goto action (\ref{eq:2-1}) is reproduced \footnote{The action of a relativistic massive point particle is $S=-m\int d\tau \sqrt{-\dot{x}^{\mu }\dot{x}_{\mu } }$. This can be written as $S=\int d\tau  \left[ \dot{x}^{\mu }p_{\mu } -\frac{1}{2}v\left( p^{2}+m^{2} \right) \right]$, where $p_{\mu }$ is conjugate to $x^{\mu }$ and $v$ is the Lagrange multiplier for the off-shell condition, $p^2 +m^2 =0$. The former action is reproduced by the eliminating $p_{\mu }$ and $v$ in order. }.

The new action (\ref{eq:2-2}) has a local symmetry under the transformations
{\setlength\arraycolsep{1pt}\begin{eqnarray}\begin{split}
\delta \mathbf{X}^{\mu } &= \alpha \mathbf{P}^{\mu } +\beta \mathbf{X}'^{\mu } , \\
\delta \mathbf{P}_{\mu } &= T^2 (\alpha \mathbf{X}'_{\mu })' +(\beta \mathbf{P}_{\mu })' , \\
\delta V &= \dot{\alpha } +U'\alpha -U\alpha ' +V'\beta -V\beta ' , \\
\delta U &= \dot{\beta } +U'\beta -U\beta ' +T^2 (\alpha V'-\alpha 'V) ,
\end{split}\label{eq:2-3}
\end{eqnarray}}
where $\alpha (\tau , \sigma )$ and $\beta (\tau , \sigma )$ are arbitrary functions.
We choose light-cone gauge to fix this local symmetry and investigate whether there is Lorentz anomaly or not. Then we will find that "anyons" appear in the spectrum \cite{bib:003}.

From now on, we set $T$ to be zero to investigate the tensionless string. 
The action (\ref{eq:2-2}) with $T=0$ is 
\begin{eqnarray}
S[\mathbf{X},\mathbf{P};V,U] = \int d\tau \oint \frac{d\sigma }{2\pi } \left\{ \dot{\mathbf{X}} ^{\mu } \mathbf{P}_{\mu } -\frac{1}{2} V \mathbf{P}^2  -U \mathbf{X}'^{\mu }\mathbf{P}_{\mu } \right\} ,
\label{eq:2-4}
\end{eqnarray}
and the gauge symmetry (\ref{eq:2-3}) becomes 
{\setlength\arraycolsep{1pt}\begin{eqnarray}\begin{split}
\delta \mathbf{X}^{\mu } &= \alpha \mathbf{P}^{\mu } +\beta \mathbf{X}'^{\mu } , \\
\delta \mathbf{P}_{\mu } &= (\beta \mathbf{P}_{\mu })' , \\
\delta V &= \dot{\alpha } +U'\alpha -U\alpha ' +V'\beta -V\beta ' , \\
\delta U &= \dot{\beta } +U'\beta -U\beta ' . 
\end{split}\label{eq:2-5}
\end{eqnarray}}
In the next subsection we fix the gauge symmetry (\ref{eq:2-6}) with light-cone gauge.

\subsection{Light-cone gauge}
The light-cone components of coordinates and their conjugates , $(X^{+},X^{-},X)$ and $(P_{+},P_{-},P)$, are written with the components in Minkowski base as follows:
\begin{eqnarray}\begin{split}
& X^{\pm } \equiv \frac{1}{\sqrt{2}} (\mathbf{X}^1\pm \mathbf{X}^0), \ X\equiv \mathbf{X}^2, \\
& P_{\pm } \equiv \frac{1}{\sqrt{2}} (\mathbf{P}_1\pm \mathbf{P}_0)=P^{\mp }, \ P\equiv \mathbf{P}_2 . 
\end{split}\label{eq:2-6}
\end{eqnarray}
We impose light-cone gauge to fix the gauge symmetry (\ref{eq:2-5}),
\begin{eqnarray}
X^{+} =\tau , \ P_{-}=p_{-}(\tau ) \not= 0 ,
\label{eq:2-7}
\end{eqnarray}
where $p_{-}(\tau )$ is a non-vanishing function of $\tau $.
This gauge choice restricts gauge parameters such that $\alpha =0, \beta =\beta _{0}(\tau )$ and then leaves only the residual global gauge symmetry induced by a constant shift of $\sigma $. 
We leave this for a moment to clarify what the constraint is, though we will fix this later.

To obtain the action in light-cone gauge, we decouple the center of mass coordinate which is the average about $\sigma $ from the rest.
Namely, for a given function $F(\tau ,\sigma )$, we decompose it into $f$ and $\bar{F} $:
\begin{eqnarray}\begin{split}
f(\tau ) \equiv  \oint \frac{d\sigma }{2\pi } F(\tau ,\sigma ) , \\
\bar{F} (\tau ,\sigma ) \equiv F(\tau ,\sigma ) -f(\tau ) .
\end{split}\label{eq:2-8}
\end{eqnarray}
Note that $\oint \frac{d\sigma }{2\pi } \bar{F}=0 $.

Using the gauge choice (\ref{eq:2-7}) and the decoupling with regard to $F=X^{-},X,P_{+},P,U$, we find that the Lagrangian (\ref{eq:2-4}) reduces to 
{\setlength\arraycolsep{1pt}\begin{eqnarray}\begin{split}
L &= \dot{x} p +\dot{x}^{-} p_{-} +p_{+}+\oint \frac{d\sigma }{2\pi } \dot{\bar{X} }\bar{P} -u\oint \frac{d\sigma }{2\pi } \bar{X}'\bar{P} - \oint \frac{d\sigma }{2\pi } \bar{U} \bar{X}' P \\
& +p_{-} \oint \frac{d\sigma }{2\pi } \left\{ \bar{X}^{-}\bar{U}' -V\left( P_{+} +\frac{1}{2p_{-}} P^2  \right)  \right\} ,
\end{split}\label{eq:2-9}
\end{eqnarray}}
where $\bar{X}^{-} $ is a lagarange multiplier giving the constraint $\bar{U}'=0 $. Together with $\oint \frac{d\sigma }{2\pi } \bar{U}=0 $, we obtain $\bar{U} =0$. 
On the contrary the variation of $\bar{U} $ induces the relation
\begin{eqnarray}
p_{-}(\bar{X}^{-} )'=-\bar{X}'P+\oint \frac{d\sigma }{2\pi } \bar{X}'\bar{P} .
\label{eq:2-10}
\end{eqnarray}
which we use to determine $\bar{X}^{-} $.

Moreover the variation of $V$ leads to
\begin{eqnarray}
P_{+} =-\frac{1}{2p_{-}} P^2 .
\label{eq:2-11}
\end{eqnarray}
We regard this equation as expressing $P_{+}$ in terms of other variables.
The center part of $P_{+}$ is the Hamiltonian
\begin{eqnarray}
H\equiv -p_{+} =\frac{1}{2p_{-}} \left( p^2 + \mathcal{M} ^2 \right) ,
\label{eq:2-12}
\end{eqnarray}
and the mass squared is given by
\begin{eqnarray}
\mathcal{M} ^2 =2p_{+}p_{-}-p^2 = \oint \frac{d\sigma }{2\pi } \bar{P} ^2 .
\label{eq:2-13}
\end{eqnarray}
In summary the Lagrangian reduces to
\begin{eqnarray}
L = \dot{x} p +\dot{x}^{-} p_{-} +\oint \frac{d\sigma }{2\pi } \dot{\bar{X} }\bar{P} -H -u\oint \frac{d\sigma }{2\pi } \bar{X}'\bar{P} .
\label{eq:2-14}
\end{eqnarray}

Next we use the residual gauge symmetry induced by $\beta _{0}$ to fix $u=0$ and rewrite the Lagrangian to the form 
\begin{eqnarray}
L = \dot{x} p +\dot{x}^{-} p_{-} +\oint \frac{d\sigma }{2\pi } \dot{\bar{X} }\bar{P} -H 
\label{eq:2-15}
\end{eqnarray}
with a constraint
\begin{eqnarray}
\oint \frac{d\sigma }{2\pi } \bar{X}'P =0 .
\label{eq:2-16}
\end{eqnarray}

Finally we solve most of the infinite constrains in the action (\ref{eq:2-4}) by eq. (\ref{eq:2-10}) and (\ref{eq:2-11}) and then leave the only constraint (\ref{eq:2-16}).
It is an advantage of using light-cone gauge that we can solve most of the constraints leaving a finite number of simple constarints.
The difference between the light-cone quantization and others originates in whether the number of constraints is finite or infinite and whether we can deal with them adequately. 

\subsubsection*{Fourier expansion}
To solve eq. (\ref{eq:2-10}) and (\ref{eq:2-11}) explicitly, we use the Fourier expansion of $X$ and $P$ with respect to $\sigma $:
{\setlength\arraycolsep{1pt}\begin{eqnarray}\begin{split}
X &= \sum _{n=-\infty }^{\infty } X_{n} e^{in\sigma } , ~~ X_{0}=x, \\
P &= \sum _{n=-\infty }^{\infty } P_{n} e^{in\sigma } , ~~ P_{0}=p.
\end{split}\label{eq:2-17}
\end{eqnarray}}
The reality conditions of $X$ and $P$ lead to 
{\setlength\arraycolsep{1pt}\begin{eqnarray}
&& (X_{n})^{*}=X_{-n}, \nonumber \\
&& (P_{n})^{*}=P_{-n},  
\label{eq:2-17-2}
\end{eqnarray}}
where the asterisk represents the complex conjugate.
From now on the sum without an explicit range specified must be understood to run from minus infinity to infinity \footnote{For example, $\sum _{n} \equiv \sum _{n=-\infty }^{\infty }$, $\sum _{n\not= 0} \equiv \sum _{n=-\infty }^{-1}+\sum _{n=1}^{\infty }$, $\sum _{n>0} \equiv \sum _{n=1}^{\infty }$ and so on.}. 
For a tensile string, we usually combine $X_{n}$ and $P_{n}$ as 
\begin{eqnarray}\begin{split}
\alpha _{n} &= -i \sqrt{\frac{T}{2}} n X_{n} +\frac{1}{\sqrt{2T}}P_{n}, \\
\tilde{\alpha }_{-n} &= i \sqrt{\frac{T}{2}} n X_{n} +\frac{1}{\sqrt{2T}}P_{n}
\end{split}\label{eq:2-18}
\end{eqnarray}
which express the right-moving or the left-moving respectively. 
However, because we have no scale like $T$, it is not clear whether we should introduce some scale to combine Fourier coefficient (\ref{eq:2-17}) in the oscillator form.

Let us rewrite the mass squared (\ref{eq:2-12}) and the constraint (\ref{eq:2-16}). 
First we solve eq. (\ref{eq:2-10}) as follows \footnote{The condition $\oint \frac{d\sigma }{2\pi } \bar{X}^{-} =0$ leads to the same constraint as (\ref{eq:2-16}) or (\ref{eq:2-24}).}: 
\begin{eqnarray}
\bar{X}^{-} = - \frac{1}{p_{-}} \sum _{n\not= 0} \frac{i}{n} M_{n} e^{in\sigma } ,
\label{eq:2-19}
\end{eqnarray}
where
\begin{eqnarray}
M_{n} \equiv -i \sum _{m} mX_{m}P_{n-m} , ~~~ \mbox{for} ~~ n\not= 0.
\label{eq:2-20}
\end{eqnarray}
Denoting the center part by $x^{-}$, we have $X^{-}=x^{-}+\bar{X}^{-} $.

Next we solve eq. (\ref{eq:2-11})
\begin{eqnarray}
P_{+} = -\frac{1}{2p_{-}} \sum _{n} L_{n} e^{in\sigma } ,
\label{eq:2-21}
\end{eqnarray}
where 
\begin{eqnarray}
L_{n} \equiv \sum _{m} P_{m}P_{n-m}.
\label{eq:2-22}
\end{eqnarray}
From the zero mode $L_{0}$, the mass squared reads as 
\begin{eqnarray}
\mathcal{M}^2 = 2\sum _{n>0} P_{n}P_{-n}.
\label{eq:2-23}
\end{eqnarray}
Further the constraint (\ref{eq:2-16}) is expressed as
\begin{eqnarray}
0 = \oint \frac{d\sigma }{2\pi } \bar{X}'P = i \sum _{n} nX_{n}P_{-n} \equiv -M_{0},
\label{eq:2-24}
\end{eqnarray}
which corresponds to the level-matching condition for a tensile string.

\subsubsection*{Equations of motion}
Using the Fourier expansion, the Lagrangian (\ref{eq:2-15}) is written in the form 
\begin{eqnarray}
L = \dot{x}p+\dot{x}^{-}p_{-}+\sum _{n\not= 0} \dot{X}_{n}P_{-n} -H , 
\label{eq:2-25}
\end{eqnarray}
where $H=\frac{1}{2p_{-}}(p^2+\mathcal{M}^2)$ and the mass squared is given by eq. (\ref{eq:2-24}).
We obtain the equations of motion from this Lagrangian, 
\begin{eqnarray}
\dot{p}=\dot{p}_{-}=0, \ \dot{x}=\frac{p}{p_{-}} , \ \dot{x}^{-}=-\frac{H}{p_{-}}
\label{eq:2-26}
\end{eqnarray}
and
\begin{eqnarray}
\dot{X}_{n}=\frac{P_{n}}{p_{-}}, ~~ \dot{P}_{n}=0.     
\label{eq:2-27}
\end{eqnarray}
These equations indicate that the center part moves with uniform velocity and the shape of string changes in proportion to $P(\sigma )$, 
\begin{eqnarray}
X(\tau ,\sigma ) = X(\tau =0,\sigma ) +\frac{P(\sigma )}{p_{-}}\tau ,
\label{eq:2-28}
\end{eqnarray}
which is expected from the equations of motion before taking the Fourier expansion.
It reflects the fact that the Hamiltonian and the mass squared are independent of $\tau $.

\section{Quantization of a 3D tensionless string and Space-time conformal symmetry}
In this section we first quantize a 3D tensionless string represented in section 2 and next find that this theory has no anomaly for the space-time conformal symmetry.

\subsection{Quantization}
In the quantum theory, the canonical variables in the action (\ref{eq:2-15}) are promoted to operators with the commutation relations
\begin{eqnarray}
[x^{-},p_{-}]=i,\ [X(\sigma ),P(\sigma ')]=2\pi i \delta (\sigma -\sigma ')  \ \ \ \mbox{with others vanishing},
\label{eq:3-1}
\end{eqnarray}
where we set $\hbar =1$. In the mode expansion, the last relation indicates
\begin{eqnarray}
[x,p]=i, \ [X_{n},P_{m}]=i\delta _{n+m,0} ,
\label{eq:3-2}
\end{eqnarray}
where $n,m\in \mathbb{Z}$.
Using these basic relations, we obtain that $L_{n}$ and $M_{n}$ satisfy the following relations:
\begin{eqnarray}\begin{split}
& [X_{n}, M_{m}]=(n+m)X_{n+m} ,~~ [P_{n},M_{m}]=nP_{n+m} ,~~ [X_{n}, L_{m}]=2iP_{n+m} ,~~ [P_{n},L_{m}]=0 \\
& [M_{n},M_{m}]=(n-m)M_{n+m} , ~~ [L_{n},M_{m}]=(n-m)L_{n+m} , ~~ [L_{n}, L_{m}]=0 ,
\end{split}\label{eq:3-3}
\end{eqnarray}
where we use the operator-ordering given in eq. (\ref{eq:2-20}) and (\ref{eq:2-22}) .
We make a remark that $L_{n}$ and $M_{n}$ satisfy  the 2-dimensional Galilean Conformal Algebra (2d GCA) \footnote{Recently this algebra was investigated in terms of a tensionless string \cite{bib:012}}.

The quantum Hamiltonian and the mass suared are then 
\begin{eqnarray}\begin{split}
& H=\frac{1}{2p_{-}}(p^2+\mathcal{M}^2), \\
& \mathcal{M}^2=2\sum _{n>0} P_{n}P_{-n}, 
\end{split}\label{eq:3-4}
\end{eqnarray}
where there is no constant term arising from ambiguity of the operator ordering because $P_{n}$ commute with $P_{-n}$ \footnote{In the case of a tensile string, the mass squared has a constant $a$ arising from the operator ordering ambiguity. 
To avoid the Lorentz anomaly, we choose the critical dimension of the string theory and the ordering constant to be $D=26$ and $a=1$, respectively. 
However in three dimensions no Lorentz anomaly exists trivially and then $a$ remains as an arbitrary constant \cite{bib:003}.}. 
On the contrary, the constraint (\ref{eq:2-24}) has ambiguity of the operator ordering. 
This ambiguity is related to the choice of the vacuum. 
Here we define $M_{0}$ as in (\ref{eq:2-24}) in order that the action of $M_{0}$ on physical state vanishes. 

\subsection{Generators }
In light-cone gauge quantization, the space-time Lorentz and conformal symmetries are not clear. 
Therefore we check whether the generators of these symmetries satisfy the expected commutation relations (\ref{eq:a2-7}) and (\ref{eq:a2-11}), respectively. 
In $D>3$, we determine the critical dimension of a bosonic string and the ordering constant in the mass squared to preserve the Lorentz invariance in quantum theory. 
In three dimensions, there is only one transverse direction and the dangerous commutator (\ref{eq:1-1}) vanishes trivially. Hence we have no Lorentz anomaly. 
However the conformal symmetry is not trivial even in three dimensions. We now investigate whether a 3D tensionless closed string has conformal symmetry.

The conformal group is generated by translations, Lorentz rotations, dilatation and specail conformal transformations.
Now we define these generators in the"reference order," which all $P_{n}$ and $p_{-}$ are to the right of $X_{n}$ and $x^{-}$ respectively \cite{bib:007}. We shall call the reference order "R-order" \footnote{We can obtain the R-ordering from the normal ordering in the tensionless limit $T\rightarrow 0$. In detail, the string ground state $|0\rangle _{T}$ of a tensile string with a tension $T$ is annihilated by positive modes of right-moving and left-moving oscillators, $\{ \alpha _{n}, ~ \tilde{\alpha }_{n} ; n>0 \} $. According to eq. (\ref{eq:2-18}), this string ground state in the tensionless limit reduces the vacumm which annihilates all $P_{n}$ for all non-zero $n$. However, when we set $T=0$ from the beginning, there is no reason why this state should be chosen. }. 
The definitions of $L_{n}$ and $M_{n}$ in eq. (\ref{eq:2-20}) and (\ref{eq:2-22}) were already into the R-order.
Hereafter, when we want to emphasize operators to be into the R-order, we specify the R-ordered operator by the subscript $R$.

We first define the translation generators as 
\begin{eqnarray}
\mathcal{P}_{R}^{\mu } \equiv \oint \frac{d\sigma }{2\pi } P^{\mu }.
\label{eq:3-5}
\end{eqnarray}
In the light-cone base, these are
\begin{eqnarray}
\mathcal{P}_{R} = p, ~~~ \mathcal{P}_{R}^{+}=p_{-},~~~\mathcal{P}_{R}^{-}=p_{+}=-H.
\label{eq:3-6}
\end{eqnarray}
Second, the Lorentz generators are defined as
\begin{eqnarray}
\mathcal{J}_{R}^{\mu } \equiv \epsilon ^{\mu \nu \rho } \oint \frac{d\sigma }{2\pi } X_{\nu }P_{\rho } .
\label{eq:3-7}
\end{eqnarray}
In light-cone base, these are written as
\begin{eqnarray}\begin{split}
& \mathcal{J}_{R} =x^{-}p_{-}+\tau H,~~~\mathcal{J}_{R}^{+}=\tau p-xp_{-}, \\
& \mathcal{J}_{R}^{-} =-x^{-}p -xH +\frac{\Lambda }{p_{-}},
\end{split}\label{eq:3-8}
\end{eqnarray}
where 
\begin{eqnarray}
\Lambda = p_{-} \oint \frac{d\sigma }{2\pi } \left[ \bar{X}\bar{P}_{+} -\bar{X}^{-}\bar{P} \right] =\sum _{n\not= 0} \left( -\frac{1}{2} X_{n}L_{-n} +\frac{i}{n} M_{n}P_{-n} \right) .
\label{eq:3-9}
\end{eqnarray}
Next the dilatation generator is defined as 
\begin{eqnarray}
\mathcal{D}_{R} = \oint \frac{d\sigma }{2\pi } X^{\mu }P_{\mu } ,
\label{eq:3-10}
\end{eqnarray}
and now expressed as
\begin{eqnarray}
\mathcal{D}_{R} =x^{-}p_{-}-\tau H +\sum _{n} X_{n}P_{-n}.
\label{eq:3-11}
\end{eqnarray}
At last the generators of the special conformal transformations are defined as 
\begin{eqnarray}
\mathcal{K}_{R}^{\mu } = \oint \frac{d\sigma }{2\pi } \left[ X^{\mu } \left( X \cdot P \right) - \frac{1}{2} \left( X \cdot X \right) P^{\mu } \right] _{R} ,
\label{eq:3-12}
\end{eqnarray}
where the subscript $R$ in the right-hand side indicates the reordering into R-order \footnote{Because $\mathcal{K}_{R}^{-}$ includes a quadratic term of $X^{-}$, the simply defined $\mathcal{K}^{-}$ is not into R-order as it is.}. 
In light-cone base, 
{\setlength\arraycolsep{1pt}\begin{eqnarray}\begin{split}
\mathcal{K}_{R}^{+} &= -\frac{1}{2}\sum _{n} X_{n}X_{-n} p_{-} + \tau \sum _{n} X_{n}P_{-n} -\tau ^2 H \\
\mathcal{K}_{R} &= xx^{-}p_{-}+\sum _{n\not= 0} \frac{i}{n}X_{n}M_{-n} +\frac{1}{2}\sum _{n} \sum _{m} X_{n}X_{m}P_{-n-m} +\tau \mathcal{J}^{-} \\
\mathcal{K}_{R}^{-} &= x^{-}x^{-}p_{-} +x^{-}\sum _{n} X_{n}P_{-n} -\frac{i}{p_{-}} \sum _{n}\sum _{m\not= 0}\left( \frac{n}{m^2}+\frac{1}{m} \right) X_{n}M_{m}P_{-n-m} \\
&  +\frac{1}{4p_{-}}\sum _{n}\sum _{m} X_{n}X_{m}L_{-n-m} .
\end{split}\label{eq:3-13}
\end{eqnarray}}

Physical observables should be represented by Hermitian operators. We require the conformal generators to be Hermite.
In our case the R-ordered definitions of the generators in (\ref{eq:3-8})(\ref{eq:3-11})(\ref{eq:3-13}) are simple but not Hermite \footnote{Of course the translation $\mathcal{P}=p$, the Hamiltonian $H$ and the mass squared $\mathcal{M}^2$ are clearly Hermite. Moreover $\Lambda $ is Hermite, and if $\sum _{n}n = 0$ the constraint $M_{0}$ is also Hermite.}.  
Then we introduce the Hermitian version $\mathcal{G}$ of the R-ordered generator $\mathcal{G}_{R}$ as follows:
\begin{eqnarray}
\mathcal{G} \equiv \frac{1}{2}\left( \mathcal{G}_{R} + (\mathcal{G}_{R} ) ^{\dagger } \right) .
\label{eq:3-14}
\end{eqnarray}
All Hermitian versions of generators are independent of $\tau $. 
Hence, when we deal with Hermitian generators, we can use generators to be set $\tau =0$. 
Here note that the differences of the Hermitian version from the R-ordered generators mostly include various divergent terms.
So, if we order generators and commutators into R-order, we need some regularization.

\subsection{Anomaly}
In the previous subsection, we defined the generators of the space-time conformal symmetry.
In three dimensions, there is the only one transverse direction \footnote{So we omit the label $I$ or $J$. $\mathcal{P}^{I}\rightarrow \mathcal{P}$, $\mathcal{K}^{I}\rightarrow \mathcal{K}$, $\mathcal{J}^{\pm I}\rightarrow \mathcal{J}^{\pm }$. And we define $\mathcal{J}^{+-}\rightarrow -\mathcal{J}$.} and the dangerous commutator (\ref{eq:1-1}) vanishes trivially. Therefore all commutation relations of Poincar\'{e} group (\ref{eq:a2-7}) are satisfied. 
In this subsection, we investigate whether the anomaly arises in the quantization of a 3D tensionless string, namely whether all commutation relations of conformal symmetry (\ref{eq:a2-12}) are satisfied. 

In \cite{bib:007}, J. Isberg et al. showed that in $D>3$ anomalies arise from the commutators, $[\mathcal{K}^{I}, \mathcal{J}^{-J}] $, $[\mathcal{K}^{+}, \mathcal{K}^{-}] $, $[\mathcal{K}^{I}, \mathcal{K}^{-}] $ and $[\mathcal{K}^{-}, \mathcal{J}^{-I}] $. 
The first commutator of these has the traceless part $L^{IJ}_{t.l.}$ with respect to $I,J$ as well as the trace part. 
The difference of trace part can be absorbed in the redefinition of $\mathcal{K}^{-}$, but the traceless part $L^{IJ}_{t.l.}$ remains as the anomaly.

In three dimensions, this commutator is only one and hence this type of anomaly does not exist.
But we need to calculate other non-trivial commutators to check the space-time conformal symmetry.
In three dimensions, $[\mathcal{K}, \mathcal{J}^{-}]$ corresponds to $[\mathcal{K}^{I}, \mathcal{J}^{-J}] $ in $D>3$. 
Though this is slightly different from $\mathcal{K}^{-}=\frac{1}{2}(\mathcal{K}^{-}_{R}+(\mathcal{K}^{-}_{R})^{\dagger })$, we can interpret that this give the redefinition as
\begin{eqnarray}
\hat{\mathcal{K} }^{-} =i[\mathcal{K}, \mathcal{J}^{-}] =\mathcal{K}^{-}+\delta \mathcal{K}^{-} ,
\label{eq:3-15}
\end{eqnarray}
where $\delta \mathcal{K}^{-}$ is a constant times $\frac{1}{p_{-}}$.
Note that in the term with $M_{0}$, we put $M_{0}$ to the right of other operators and set zero on physical Hilbert space if we need.

We still have the three "dangerous" commutators.
Using the Jacobi identity and $[\mathcal{K}, \hat{\mathcal{K}}^{-}]=0$ which has no anomaly, we find that $[\mathcal{K}^{+}, \hat{\mathcal{K}}^{-}]$ has no anomaly, that is, $[\mathcal{K}^{+}, \hat{\mathcal{K}}^{-}]=0$.
On the other hand, it is non-trivial whether the last two "dangerous" commutators vanish and then we have to check it. 
Fortunately, because there are the relations between $[\mathcal{K}^{I}, \mathcal{K}^{-}] $ and $[\mathcal{K}^{-}, \mathcal{J}^{-I}] $\footnote{$[\mathcal{K}, \hat{\mathcal{K}}^{-}]=-i[[\mathcal{J}^{-}, \hat{\mathcal{K}}^{-}], \mathcal{K}^{+}]$ and $[\hat{\mathcal{K}}^{-}, \mathcal{J}^{-}]=-i[\mathcal{P}^{-}, [\hat{\mathcal{K}}^{-}, \mathcal{K}]]$.}, all we have to do is the calculation of one commutator. 
We calculate the easier one, $[\mathcal{K}^{-}, \mathcal{J}^{-I}] $.

$[\hat{\mathcal{K}}^{-}, \mathcal{J}^{-} ] $ must be zero:
\begin{eqnarray}
[\hat{\mathcal{K}}^{-}, \mathcal{J}^{-} ] =0 .
\label{eq:3-16}
\end{eqnarray}
This commutator can have parts which are in proportion to $(x^{-})^2$ , $x^{-}\frac{1}{p_{-}}$ and $\frac{1}{p_{-}{}^2}$ in the R-ordering and complicated divergences too. 
In particular, the part in proportion to $\frac{1}{p_{-}{}^2}$ has quintic, qubic and linear terms with respect to $X_{n}$ and $P_{n}$ and is complicated.
So the calculation of this commutator needs a lot of labor and care.

Though the calculations of other commutators also have divergences, we can deal with them without the concrete regularization. 
However, commutators (\ref{eq:3-15}) (\ref{eq:3-16}) have many types of divergences \footnote{For example, $\sum _{n}1$ and $\sum _{n} n$.} and it is too complicated to calculate them correctly. 
Moreover we should take care of the shift of dummy variables in the sum and the termwise re-summation\footnote{For operators into some order, $\sum _{n}X_{n}P_{-n}|_{*} = \sum _{n}X_{n+k}P_{-n-k}|_{*} $, where the subscripts mean these terms are into some order. But for the number, $\sum _{n}n \not= \sum _{n}(n+k) \not= \sum _{n}n +k\sum _{n}1$. }.
Then the regularization help us from these divergences.

\subsubsection*{Cutoff regulerization}
We use the cutoff regularization to remove higher modes of $X$ and $P$, 
\begin{eqnarray}
X_{n}=P_{n}=0 ~~~\mbox{for} ~~ |n|>N, 
\label{eq:3-17}
\end{eqnarray}
where we assume that $N$ is a large integer. 
Using this, we obtain
\begin{eqnarray}
M_{n}=-i \sum _{|m|\leq N, |n-m|\leq N} mX_{m}P_{n-m} , ~~~ L_{n}=\sum _{|m|\leq N, |n-m|\leq N} P_{m}P_{n-m} ,
\label{eq:3-18}
\end{eqnarray}
and also find $M_{n}=L_{n}=0$ for $|n|>2N$.

Thanks to the cutoff regularization, summations reduce to finite sum and term-wise treatments are possible. 
And the error as to the shift of dummy variables will decrease.
Further, because of the symmetric cutoff with respect to positive and negative modes, we get $\sum _{n}n = \sum _{n} \frac{1}{n} =0$.
Thus we resolve many difficulties for the divergence.
But we have to take care of the informations about the range of the summation.

Then there remains only the lengthy calculation for us.
After the calculation under the cutoff regularization, we take the limit $N\rightarrow \infty $.

\subsection{Check of anomaly free}
Our goal is to check a relation
\begin{eqnarray}
[\hat{\mathcal{K}}^{-}, \mathcal{J}^{-}]=0
\label{eq:3-19}
\end{eqnarray}
under the cutoff regularization. 
We can check that most of other commutation relations are satisfied without using the cutoff regularization and that the rest of commutation relations are satisfied by using eq. (\ref{eq:3-19}). 

We summary steps of the calculation, instead of representing the process of the lengthy calculation explicitly. 
We can deform eq.(\ref{eq:3-19}) to
\begin{eqnarray}
[\hat{\mathcal{K}}^{-}, \mathcal{J}^{-}] = \left[ \frac{1}{2}\left( \mathcal{K}^{-}_{R} +(\mathcal{K}^{-}_{R})^{\dagger } \right) +\delta \mathcal{K}^{-}, \mathcal{J}^{-} \right] =\frac{1}{2} \left( [\mathcal{K}_{R}^{-}, \mathcal{J}^{-} ] -[\mathcal{K}_{R}^{-}, \mathcal{J}^{-} ] ^{\dagger } \right) +[\delta \mathcal{K}^{-}, \mathcal{J}^{-}] .
\label{eq:3-20}
\end{eqnarray}
If we use Hermitian operators from the beginning, we have more terms than in the case of R-ordered operators and many divergent terms.
Then we first use R-ordered operator, $\mathcal{J}^{-}_{R}$, and next consider its Hermite-conjugate and finally add the commutator of the difference $\delta \mathcal{K}^{-}$.

First we calculate 
\begin{eqnarray}
[\mathcal{K}_{R}^{-}, \mathcal{J}^{-} ] ,
\label{eq:3-21}
\end{eqnarray}
and order the result into the R-order except for terms with $M_{0}$, which is the right of all other operators.
Here note that $\mathcal{J}^{-} = \frac{1}{2} \left( \mathcal{J}^{-}+(\mathcal{J}^{-})^{\dagger } \right) = \mathcal{J}_{R}^{-} +\frac{1}{2}i\frac{p}{p_{-}}$.
The calculation of eq. (\ref{eq:3-21}) is very lengthy and laborious. 
The commutator (\ref{eq:3-21}) can have three parts which are proportional to $(x^{-})^2$, $x^{-}\frac{1}{p_{-}}$ and $\frac{1}{p_{-}{}^{2}}$.
We can check the vanishing of parts proportional to $(x^{-})^2$ and $x^{-}\frac{1}{p_{-}}$ easily.
The part proportional to $\frac{1}{p_{-}{}^2}$ has quintic terms with $XXPPP$-form , cubic terms with $XPP$-form and linear terms with $P$-form. 
Maximal quintic terms are independent of the ordering. Therefore we get the same result as the classical case.
In fact, after the lengthy calculation, we find the absence of quintic terms.
However cubic terms and linear terms do not vanish. So we must consider other parts of (\ref{eq:3-20}).

Next we consider the Hermite-conjugate of (\ref{eq:3-21}) and then make the anti-Hermitian version of (\ref{eq:3-21}).
In the anti-Hermitian version (\ref{eq:3-21}), we can deform all cubic terms to the form with $[X,P]$ because all summations have the symmetry which all dummy variables invert simultaneously\footnote{For example, its anti-Hermitian version of $\sum _{|n|\leq N}\sum _{0<|m|\leq N} \frac{n}{m} X_{n}P_{m}P_{-n-m}$ is \\
$\frac{1}{2} \sum _{|n|\leq N}\sum _{0<|m|\leq N} \frac{n}{m} (X_{n}P_{m}P_{-n-m} -P_{n+m}P_{-m}X_{-n}) = \frac{1}{2} \sum _{|n|\leq N}\sum _{0<|m|\leq N} \frac{n}{m} (X_{n}P_{m}P_{-n-m} -P_{-n-m}P_{m}X_{n}) = \frac{1}{2} \sum _{|n|\leq N}\sum _{0<|m|\leq N} \frac{n}{m} [X_{n},P_{m}] P_{-n-m} $. In the case of quintic terms or linear terms, we will get anti-commutator. }. 
Therefore all cubic terms reduce to the linear term.
There is this symmetry in the R-ordering, but not always in other orderings, like normal ordering.

In this way, all we have to check is the linear terms.
The sum of subscripts in each terms is zero and hence the linear terms with $P$-form must be the form that a constant times $P_{0}=p$.
Because the calculation of only linear terms is the same as calculating cubic terms, we calculate all terms with $p$.  
Though we have more hard work apparently, the calculation becomes easier because we don't have to distinguish the degree of operators.
The part with $p$ of (\ref{eq:3-21}) is \footnote{If we use the following equations, the calculation may become easier:\\
$[\mathcal{K}_{R}^{-}, \mathcal{J}^{-} ] |_{p} = \left( [\mathcal{K}_{R}^{-}|_{p}, \mathcal{J}^{-} ] +[\mathcal{K}_{R}^{-}, \mathcal{J}^{-}|_{p} ] -[\mathcal{K}_{R}^{-}|_{p}, \mathcal{J}^{-}|_{p} ] \right) |_{p}$ and $\mathcal{K}^{-}_{R}|_{p}=-\mathcal{K}_{R}\frac{p}{p_{-}}+\frac{1}{2}\mathcal{K}^{+}\frac{p^2}{p_{-}{}^2}+2x^{-}xp +x\sum _{n} X_{n}P_{-n} \frac{p}{p_{-}}$, $[\mathcal{K}_{R}, \mathcal{J}^{-}] = -i(\mathcal{K}^{-}_{R}+\delta \mathcal{K}^{-}_{R})$ and $[\mathcal{K}^{+}, \mathcal{J}^{-}] =i\mathcal{K}$.}
\begin{eqnarray}
[\mathcal{K}_{R}^{-}, \mathcal{J}^{-} ] |_{p} = \frac{1}{2}i\delta \mathcal{K} \frac{p^2}{p_{-}^{2}} +i\delta \mathcal{K}^{-}_{R} \frac{p}{p_{-}},
\label{eq:3-22}
\end{eqnarray}
where $(\cdots )|_{p}$ represents the part with $p$ of $(\cdots )$ and we define $\delta \mathcal{K} \equiv \mathcal{K}-\mathcal{K}_{R} = x \times (\mbox{constant})$ and $\delta \mathcal{K}^{-}_{R} \equiv i[\mathcal{K}_{R}, \mathcal{J}^{-}] -\mathcal{K}^{-}_{R}$ \footnote{In the representation without regularization, $\delta \mathcal{K} = -\frac{1}{2}i(1+\sum _{n}1)x$ and\\
$\delta \mathcal{K}^{-}_{R} =-\frac{i}{2}x^{-} +\frac{1}{2p_{-}} +\frac{i}{2}(\sum _{l\not= 0} 1)\sum _{n \not= 0} X_{n}P_{-n} \frac{1}{p_{-}} +\frac{i}{2}(\sum _{n} 1) x\frac{p}{p_{-}} +i\sum _{n \not=0} \sum _{m \not= 0} \frac{n^2}{m^2} X_{n}P_{-n} \frac{1}{p_{-}} +2i\sum _{n \not= 0} \frac{1}{n} X_{n}P_{-n} \frac{1}{p_{-}} M_{0}$.}. Furthermore we obtain 
\begin{eqnarray}
\frac{1}{2}\left( [\mathcal{K}_{R}^{-}, \mathcal{J}^{-} ] |_{p} - ([\mathcal{K}_{R}^{-}, \mathcal{J}^{-} ] |_{p})^{\dagger } \right) = i\delta \mathcal{K}^{-} \frac{p}{p_{-}} ,
\label{eq:3-23}
\end{eqnarray}
where we use $\delta \mathcal{K}^{-} =\frac{1}{2}(\delta \mathcal{K}^{-}_{R} + (\delta \mathcal{K}^{-}_{R} )^{\dagger })$.

Finally we calculate the commutator of the difference $\delta \mathcal{K}^{-}$.
\begin{eqnarray}
[\delta \mathcal{K}^{-}, \mathcal{J}^{-}] = [\delta \mathcal{K}^{-}, -x^{-}p] = -i \delta \mathcal{K}^{-} \frac{p}{p_{-}},
\label{eq:3-24}
\end{eqnarray}
where we use $\delta \mathcal{K}^{-} = \frac{1}{p_{-}} \times ( \mbox{constant} )$ and $[\frac{1}{p_{-}}, x^{-}] =i\frac{1}{p_{-}{}^2}$.

Adding (\ref{eq:3-23}) and (\ref{eq:3-24}), we can check the commutation relation (\ref{eq:3-19}).
Thus we find that there is no anomaly of the space-time conformal symmetry in a 3D tensionless bosonic closed string.

\section{Spectrum}
In this section, we investigate the spectrum of a 3D tensionless closed string.
We deal the center of mass part and the rest separately and assume that the ground state of the center is $|p,p_{-} \rangle $ like the case of a tensile string. We do not care about the ground state of the center part so much. 

On the other hand, we need the discussion about the rest part, which includes non-zero modes. 
First we consider the R-ordered vacuum $|0 \rangle _{R}$, which satisfy
\begin{eqnarray}
P_{n} |0 \rangle _{R} =0 ~~~ \mbox{for all non-zero } n ,
\label{eq:4-1}
\end{eqnarray}
as the string ground state.
We can obtain this R-ordered string ground state from the string ground state in the tensionless limit \cite{bib:007}.

When we choose this string ground state, we obtain other states by acting $\{ X_{n}; n \not= 0 \} $ on $|0 \rangle _{R}$.
The fundamental elements of states are 
\begin{eqnarray}
X_{n_1} X_{n_2} \cdots X_{n_l} |0\rangle _{R} ,
\label{eq:4-2}
\end{eqnarray}
where $n_{i} ~ (i=1,2,\cdots l ) $ are non-zero integers.
We combine these state to obtain the general states.

Here note that the physical states must satisfy $M_{0}=0$. 
Therefore the physical states are the states with the form (\ref{eq:4-2}) which satisfy
\begin{eqnarray}
\sum _{i=1}^{l}n_{i}=0 .
\label{eq:4-3}
\end{eqnarray}
The general physical states satisfy this condition in each terms.
Under this condition, we try to make eigenstates of $\mathcal{M}^2$.
In the case of the R-ordered string ground state, it is convenient to represent $P_{n}$ as 
\begin{eqnarray}
P_{n}=-i\frac{\partial }{\partial X_{-n}}, ~~~ [X_{n}, P_{m}] =i\delta _{n+m,0}
\label{eq:4-4}
\end{eqnarray}
Using this representation, $\mathcal{M}^2$ is expressed as 
\begin{eqnarray}
\mathcal{M}^2 = -2 \sum _{n>0} \frac{\partial }{\partial X_{-n}} \frac{\partial }{\partial X_{n}} .
\label{eq:4-5}
\end{eqnarray}

By the space-time conformal symmetry, we expect that eigenvalues of $\mathcal{M}^2$ are continuous or zero. 
If we find the eigenfunction with the eigenvalue $M^2$, we get an eigenfunction of $\mathcal{M}^2$ with the eigenvalue $\lambda ^{-2} M^2$ by transforming $X_{n}\rightarrow \lambda X_{n}$ for all $n$.
Here we are especially interested in the case of zero eigenvalue, that is massless.

\subsection{Mass eigenstate }
In this subsection, we use the similar method in \cite{bib:016}.
In order to find eigenvalues of $\mathcal{M}^2$ and their eigenfunctions, we rewrite eq.(\ref{eq:4-5}),
\begin{eqnarray}
\mathcal{M}^2 = - \frac{1}{2} \sum _{n>0} \left[ \frac{\partial ^2}{\partial r_{n} {}^2 } +\frac{1}{r_{n}} \frac{\partial }{\partial r_{n}} +\frac{1}{r_{n}{}^2}\frac{\partial ^2}{\partial \theta _{n}{}^2 } \right] 
\label{eq:4-6}
\end{eqnarray}
where $r_{n}$ and $\theta _{n}$ are the real operators which are defined by $X_{n}=r_{n}e^{i\theta _{n}}$ and $X_{-n}=r_{n}e^{-i\theta _{n}}$ for all positive $n$.
Because we can separate variables into each $n$, we consider the following differential equation:
\begin{eqnarray}
-\frac{1}{4} \left[ \frac{\partial ^2}{\partial r^2 } +\frac{1}{r} \frac{\partial }{\partial r} +\frac{1}{r^2}\frac{\partial ^2}{\partial \theta ^2 } \right] \psi _{m}(r, \theta ) = m^2 \psi _{m}(r, \theta ) ,
\label{eq:4-7}
\end{eqnarray}
where $r\geq 0$ and $m^2\geq 0$\footnote{Because the eigenvalue of the operator $A A^{\dagger }$ is zero or positive number, we consider the case of $m^2\geq 0$. And we also find that the solutions for $m^2 < 0$ have bad behavior in $r\rightarrow \infty $ and can not be normalized. }.
Here we assume $\psi _{m,s }(r,\theta )=\phi _{m,s }(r) e^{i s \theta }$, where $s$ is integer and $\phi _{m,s}(r)$ is the function which depends on only $r$. 
Then we obtain 
\begin{eqnarray}
\left[ \frac{d^2}{d r^2 } +\frac{1}{r} \frac{d}{d r} + 4 m^2 -\frac{s^2}{r^2} \right] \phi _{m,s }(r) = 0 .
\label{eq:4-8}
\end{eqnarray}

\subsubsection{$m^2 >0$}
For $m^2 > 0$, we replace $r$ with $\hat{r} =2m r$ and obtain
\begin{eqnarray}
\left[ \frac{d^2}{d \hat{r} ^2 } +\frac{1}{\hat{r} } \frac{d}{d \hat{r} } + 1 -\frac{s ^2}{\hat{r} ^2} \right] \phi _{m,s }\left( \frac{\hat{r} }{2m}\right) = 0 ,
\label{eq:4-9}
\end{eqnarray}
where $m > 0$. The solution of this equation is expressed in terms of Bessel function $J_{|s|}(\hat{r} )$. 
In this way, we get the solution $\psi _{m,s }(r,\theta ) = N_{m } J_{|s|}(2mr) e^{i s \theta }$, where $N_{m}$ is the normalization constant \footnote{Note that $J_{\nu }(\hat{r} )=(-1)^{\nu }J_{-\nu }(\hat{r} )$ for integer $\nu $.}.

We rewrite the solutions in terms of $X_{n}$ and $X_{-n}$ with $n>0$: 
{\setlength\arraycolsep{1pt}\begin{eqnarray}
\psi _{m,\pm |s|}(r_{n},\theta _{n}) &=& N_{m} J_{|s|}( 2m(X_{n}X_{-n})^{\frac{1}{2}}) \left( \frac{X_{n}}{X_{-n}} \right) ^{\frac{s}{2}} \nonumber \\
&=& N_{m} (m X_{\pm n} )^{|s|} \sum _{l=0}^{\infty } \frac{(-m^2)^l }{l! \ (l+|s|)!} ( X_{n}X_{-n} )^{l} ,
\label{eq:4-10}
\end{eqnarray}}
where we use 
\begin{eqnarray}
J_{\nu }(z) = \left( \frac{z}{2} \right) ^{\nu } \sum _{l=0}^{\infty } \frac{(-1)^l}{l! \ \Gamma (l+\nu +1)} \left( \frac{z}{2}\right) ^{2l}.
\label{eq:4-11}
\end{eqnarray}
Thus we find that the eigenfunctions are the combinations of positive integer power with respect to $X_{n}$ and $X_{-n}$ as in eq.(\ref{eq:4-2}).

Now we consider the normalization of wave functions. 
Since $m^2$ is not discrete valuable, the wave functions for each $m^2$ can not be normalized to "1" but the wave functions for different values of $m^2$ must be orthogonal. 
Here we consider the next scalar product. The scalar product of two wave functions $\psi _{1}(r,\theta )$ and $\psi _{2}(r,\theta )$ is
\begin{eqnarray}
(\psi _{1}, \psi _{2}) = \int _{0}^{\infty } dr \int _{0}^{2\pi } d\theta r \psi _{1}^{*}(r,\theta ) \psi _{2}(r, \theta ) .
\label{eq:4-12}
\end{eqnarray}
Then we obtain for $m>0$ and $m'>0$
\begin{eqnarray}
(\psi _{m,s },\psi _{m',s'}) = \frac{\pi }{2} \frac{|N_{m}|^2}{m} \delta (m-m') \delta _{s,s'} 
\label{eq:4-13}
\end{eqnarray}
and find the orthogonality. The detail of the normalization is given in appendix.

\subsubsection{$m^2=0$ }
For $m^2=0$, eq.(\ref{eq:4-8}) becomes
\begin{eqnarray}
\left[ \frac{d^2}{d r^2 } +\frac{1}{r} \frac{d}{d r} -\frac{s^2}{r^2} \right] \phi _{0,s}(r) = 0 .
\label{eq:4-14}
\end{eqnarray}
The solutions of this equation for $s \not= 0$ are $r^{|s|}$ and $r^{-|s|}$. For $s=0$, the solutions are constant and $\log r$.
Because of the orthogonality between two eigenfunctions with $m^2 >0$, $r^{|s|}$ for $s \not= 0$ and constant for $s=0 $ are chosen \cite{bib:016}\footnote{The detail is in appendix.}. 
In terms of $X_{n}$ and $X_{-n}$ with $n>0$, the eigenfunctions $\psi _{0,s}(r,\theta )$ are $(X_{n})^{s}$ for $s>0$, $(X_{-n})^{-s}$ for $s<0$ and constant for $s=0$.
If we collect these three cases, we get 
\begin{eqnarray}
\psi _{0,\pm |s|}(r,\theta ) \propto r^{|s|}e^{\pm i |s| \theta } = (X_{\pm n})^{|s|} .
\label{eq:4-15}
\end{eqnarray}

\subsubsection{Total eigenfunction}
The eigenfunctions of $\mathcal{M}^2$ are the product
\begin{eqnarray}
\Psi = \prod _{n>0} \psi _{m_{n},s_{n}}(r_{n},\theta _{n})
\label{eq:4-16}
\end{eqnarray}
and their eigenvalues are
\begin{eqnarray}
\mathcal{M}^2 = 2 \sum _{n>0} (m_{n})^2. 
\label{eq:4-17}
\end{eqnarray}

\subsection{Physical spectrum}
The physical state must satisfy the constraint $M_{0}=0$.
In terms of $r_{n}$ and $\theta _{n}$, $M_{0}$ becomes 
\begin{eqnarray}
M_{0}= -i\sum _{n} n X_{n}P_{-n} = -\sum _{n>0} n \left[ X_{n}\frac{\partial }{\partial X_{n}} -X_{-n}\frac{\partial }{\partial X_{-n}}\right] = i\sum _{n>0} n \frac{\partial }{\partial \theta _{n}}.
\label{eq:4-18}
\end{eqnarray}
Then the constraint in the case of (\ref{eq:4-16}) is
\begin{eqnarray}
\sum _{n>0} n s_{n} =0.
\label{eq:4-19}
\end{eqnarray}
As mentioned in the beginning of this section, the eigenvalues of $\mathcal{M}^2$ are continuous or zero.
Though this fact is expected by the conformal symmetry, it is understood by the explicit solution (\ref{eq:4-10}). 
We get continuous eigenvalues if $m_{n}^2>0$ at least for some $n$.
On the other hand, we get massless states only if $m_{n}=0$ for all $n$.
Since we are interested in massless states which are expected to preserve the space-time conformal symmetry, 
we do not discuss the continuous spectrum anymore.
Now we investigate the case of massless in detail. 

\subsubsection*{Massless states }
If we want to get massless states, we have to choose the solution (\ref{eq:4-15}) for all positive integer $n$.
In other words, we should choose the states (\ref{eq:4-2}) such that their subscripts satisfy 
\begin{eqnarray}
X_{n_1} X_{n_2} \cdots X_{n_l} |0\rangle _{R} ~~ \mbox{with } ~~ \sum _{i=1}^{l}n_{i}=0 ~~\mbox{and} ~~ n_{i}+n_{j}\not= 0 ~~ \mbox{for } {}^{\forall }i,j .
\label{eq:4-20}
\end{eqnarray}
One of the simplest examples is
\begin{eqnarray}
X_{2}X_{-1}X_{-1}|0 \rangle _{R} .
\label{eq:4-21}
\end{eqnarray}
Because a linear term of $X_{n}$ is prohibited by the constraint $\sum _{i=1}^{l}n_{i}=0$ and the squared terms like $X_{n}X_{-n}$ do not create massless states which consist of a single term: monomial-type.

By the way, there are massless states consist of not only monomial, but also polynomial. 
For example, states like $(X_{2}X_{-2}-X_{1}X_{-1})|0\rangle _{R}$ or $\left( 4(X_{2}X_{-2})(X_{1}X_{-1}) -(X_{2}X_{-2})^2 -(X_{1}X_{-1})^2\right) |0\rangle _{R}$ are also physical states with $\mathcal{M}^2 =0$.

In the 3d Poincar\'{e} group, there are two Lorentz invariant which commute each other: $\mathcal{P}^{2}=-\mathcal{M}^{2}$ and $\mathcal{P} \cdot \mathcal{J} =\Lambda $. 
The first invariant is mass squared operator and now we consider the case of $\mathcal{M}^2 =0$.
The second invariant is "spin" operator and represented as
\begin{eqnarray}
\Lambda = -\frac{1}{2} \sum _{n \not= 0} \sum _{m \not= 0, -n} \left( 1+\frac{m}{n}+\frac{n}{m} \right) X_{n+m} \frac{\partial }{\partial X_{n}}\frac{\partial }{\partial X_{m}}
\label{eq:4-22}
\end{eqnarray}
in the X-representation. We find that acting of $\Lambda $ on states decreases the number of $X$ by one.
Because of $[\mathcal{M}^2, \Lambda ]=0$, we can get some polynomial-type massless states by acting $\Lambda $ on monomial-type massless states.
For example,
{\setlength\arraycolsep{1pt}\begin{eqnarray}
\Lambda X_{2}X_{-1}X_{-1}|0\rangle _{R} &=& -3(X_{2}X_{-2}-X_{1}X_{-1})|0\rangle _{R} \nonumber \\
\Lambda X_{2}X_{2}X_{-1}X_{-1}X_{-1}X_{-1}|0\rangle _{R} &\sim & 6(2(X_{1}X_{-1})-3(X_{2}X_{-2}))X_{2}X_{-1}X_{-1}|0\rangle _{R}  \\
\Lambda \Lambda X_{2}X_{2}X_{-1}X_{-1}X_{-1}X_{-1}|0\rangle _{R} &\sim & 54((X_{1}X_{-1})^2+(X_{2}X_{-2})^2-4(X_{2}X_{-2})(X_{1}X_{-1}))|0\rangle _{R}, \nonumber 
\label{eq:4-23}
\end{eqnarray}}
where $\sim $ means the omission of monomial-type massless states.
The first line of (\ref{eq:4-23}) is the most simple polynomial-type massless state. The second and third line are polynomial-type massless states with fifth and fourth degree of $X$. 
Here note that all massless states with 3rd degree are monomial-type.

Furthermore the inner product of polynomial-type massless states and any massive states, $\psi _{m, s}$, vanish.
Because the number of the power of $r_{n}$ in polynomial-type massless states is greater than that in monomial-type massless states by even number, we will find this vanishing by using the partial integration and the reason that we use in the definition of the delta function (\ref{eq:a3-5}).

In this way, we expect that all polynomial-type massless states are created from monomial-type massless by the action of $\Lambda $ or other operators commutative with $\mathcal{M}^2$ and $M_{0}$, though we don't investigate more in this paper.

Now we consider the theory which consist only of massless states.
Such theory is expected to be space-time conformal invariant. 
Here we define the "conformal dimension"\footnote{This definition is the inverse of the natural definition of the conformal dimension.} from the dilatation as 
\begin{eqnarray}
\Delta _{R} \equiv i \mathcal{D}_{R} =\sum _{n} X_{n} \frac{\partial }{\partial X_{n}}
\label{eq:4-120}
\end{eqnarray}
and may use this operator to investigate the properties of the massless states. 
We find that $\Delta _{R}$ count the number of $X$-type operators by the commutators
\begin{eqnarray}
[\Delta _{R}, X_{n}]=X_{n} .
\label{eq:4-130}
\end{eqnarray}
For example, we can determine that the state (\ref{eq:4-20}) has $\Delta _{R}=l$ and particularly the string ground state $|0 \rangle _{R}$ has $\Delta _{R}=0$.

Though we do not investigate massless states anymore, it is interesting to characterize massless states in terms of the 3d space-time conformal symmetry.   

\section{Conclusion and Outlook}
In this paper, we showed that a 3D tensionless bosonic closed string in light-cone gauge have no anomaly of the space-time conformal symmetry under the Hermitian version of the R-ordering and the cutoff regularization. 
Further we investigated the spectrum of a 3D tensionless string, particularly massless states.
When we consider the R-ordered string ground state, we obtained the simple expression.

Note that the results we got is in the case of a single string. 
We don't understand anything about multi-string theories with interactions or string field theories, as well as in the case of \cite{bib:003, bib:004, bib:005}. 
It is interesting to investigate the effects of the interaction. 
To study this in detail, we may need the string field theory which reproduces the spectrum of a tensionless string.

A 3D tensionless string has some of prospects. 

First, one may image the open string version of our results. In the open string case, we must consider the boundary condition at two endpoints.
Even if we choose any boundary conditions, we can fix the gauge in the same way as our case.
However we need the caution about the Fourier expansion.
In the case of a tensionless (or tensile) closed string, thanks to the periodicity of $\sigma $ (or a combination of $\sigma $ and $\tau $), we can use the Fourier expansion with respect to $\sigma $ (or a combination of $\sigma $ and $\tau $). 
But in the case of a open string, the way of the expansion changes according to the boundary condition. 
If we choose Neumann boundary condition at two endpoints, the coordinate $X(\tau ,\sigma )$ can be represented with a quasi-periodic function of $\sigma$ or the combination of $\tau $ and $\sigma $.
If we choose Neumann boundary condition at one endpoint and Dirichlet boundary condition at other, $X(\tau ,\sigma )$ can be represented with a quasi-antiperiodic function of $\sigma$ or the combination of $\tau $ and $\sigma $.
If we choose Dirichlet boundary condition at two endpoints, $X(\tau ,\sigma )$ can be represented with a quasi-periodic function of $\sigma$ or the combination of $\tau $ and $\sigma $, but $X$ can have a linear term of $\sigma $.
Therefore, in a tensionless open string except for the case of the Neumann-Neumann condition, we do not know simply whether we obtain the same result as this paper. 
Furthermore note that because of the difference of the equation of motion between the tensile case and the tensionless case, the coordinate $X$ in a tensile open string are essentially periodic but in a tensionless open string is a formal expedient to represent the zero point of $X$ or its differential.
\footnote{Here we consider the case of Neumann-Neumann condition as an example. The equation of motion of a tensile string is $\left( \partial ^2_{\tau } -T^2 \partial ^2_{\sigma } \right) X(\tau ,\sigma ) =0$ and its general solution is $X(\tau ,\sigma ) = \frac{1}{2} \left( F(T\tau -\sigma ) + G(T\tau +\sigma ) \right)$. In the case of Neumann-Neumann condition such that $X'(\tau ,0)=X'(\tau ,\pi )=0$, we obtain $F'(\tau )=G'(\tau )$ and $F'(\tau +\pi )=G'(\tau -\pi )$. Using these relations, we get $G(u)=F(u)+a$ and $F(u+2\pi )=F(u)+2\pi \cdot v$ where $a$ and $v$ are constant. In this way, we can write $X(\tau ,\sigma )$ with the quasi-periodic function of the combination of $\tau $ and $\sigma $. On the other hand, the equation of motion of a tensionless string is $\partial ^2_{\tau } X(\tau ,\sigma ) =0$ and its general solution is $X(\tau ,\sigma ) = F(\sigma ) + \tau G(\sigma )$. In the case of Neumann-Neumann condition such that $X'(\tau ,0)=X'(\tau ,\pi )=0$, we get $F'(0)=F'(\pi )=0$ and $G'(0)=G'(\pi )=0$. We can choose $\cos (n\sigma ) $ with integer $n$ as the function that satisfy these relations. The case of other sets of boundary conditions at two endpoints are also discussed in the same way.}
Hence we can not use the Fourier expansion based on the periodicity simply and may need the discussion in another way. 
For example, one may deal the functions $X(\sigma )$ or $P(\sigma )$ without using the mode expansions, like the case of \cite{bib:007}.
But it may be difficult to think the concrete regularization.

Next is the relation of the other background metric, which we didn't comment on this in this paper. 
We find the invariance of the action(\ref{eq:2-4}) under the space-time Weyl transformation with the transformation of $V$. 
Then we expect applications to other background space-times, e.g. $AdS_{3}$, though we must investigate whether this theory represents a tensionless string on non-Minkowski background space-time like a point particle or whether this classical fact is satisfied in quantum theory.

As other possibilities there are the supersymmetric version, the membrane version, the relation of Higher-spin gauge theories and so on.
In the case of a tensionless closed supersymmetric string, we can use the similar way to ours, unlike a tensionless open string. 
We expect that the supersymmetric case in light-cone gauge gives the different result from the case of BRST quantization \cite{bib:018, bib:019}.
And it is interesting to consider not only string but also two dimensional object, so-called the membrane. 
Moreover we are interested in the relation of Higher-spin gauge theories \cite{bib:009,bib:010}, which may be one of the limits of the string theory.

Of course we are interested in the explanation for the difference between BRST quantization and the light-cone gauge quantization and the covariant method that reproduces results of the light-cone gauge quantization.   

\section*{Acknowledgements}
I would like to thank M. Fukuma, K. Sugiyama and S. Shimasaki for useful comments and advices.
The work of KM was supported by the Japan Society for the Promotion of Science (JSPS). 

\appendix
\section{Light-cone components}
We define the components of light-cone base in this paper. 

In D dimensions Cartesian coordinates $(\mathbb{X}^{\mu } ~;~ \mu =0,1,\cdots ,D-1)$, we define the Minkowski metric $\eta _{\mu \nu } ~;~ \mu ,\nu =0,1,\cdots , D-1$ such that 
\begin{eqnarray}
\eta =\mbox{diag}(-1,1,\cdots ,1) .
\label{eq:a1-1}
\end{eqnarray}
The light-cone components of coordinates are
\begin{eqnarray}
X^{\pm } \equiv \frac{1}{\sqrt{2}} \left( \mathbb{X}^1 \pm \mathbb{X}^0 \right) ,~~~ X^{I}\equiv \mathbb{X}^{I} ~~(I=2, \cdots ,D-1) .
\label{eq:a1-2}
\end{eqnarray}
Similarly, the light-cone components of an arbitrary vector $(\mathbb{V}_{\mu } ~;~ \mu =0,1,\cdots ,D-1)$ are 
\begin{eqnarray}
V_{\pm }\equiv \frac{1}{\sqrt{2}} \left( \mathbb{V}_1 \pm \mathbb{V}_0 \right) =V^{\mp },~~~ V_{I}\equiv \mathbb{V}_{I} ~~(I=2, \cdots ,D-1) .
\label{eq:a1-3}
\end{eqnarray}
The indices are raised and lowered with metric $\eta $. Moreover the inner product is given as 
\begin{eqnarray}
-\mathbb{V}_0^2+\sum _{i=1}^{D-1} \mathbb{V}_{i}^2 \equiv \mathbb{V}^2 = 2V_{+}V_{-} +\sum _{I=2}^{D-1}V_{I}^2
\label{eq:a1-4}
\end{eqnarray}

In three dimensions, the two rank anti-symmetric tensor is rewritten as the vector. 
For example $S^{\mu \nu }=-S^{\nu \mu }$ is represented as
\begin{eqnarray}
S^{\mu } \equiv \epsilon ^{\mu \nu \rho }S_{\nu \rho } ,
\label{eq:a1-5}
\end{eqnarray}
where $\epsilon ^{\mu \nu \rho }$ is the totally antisymmetric tensor such that $\epsilon ^{012}=1$ in Minkowski base and $\epsilon ^{+-2}=1$ in light-cone base.
Furthermore we emphasize that in three dimensions the transverse direction ($I=2, \cdots , D-1$) is only one ($I=2$).

\section{Generators}
\subsection{Poincar\'{e} group}
We expect string theory to have the Poincar\'{e} symmetry. 
The quantization can induce Lorentz anomaly, and the cancellation of this anomaly determines the critical dimension and the ordering constant. 
First we consider in general dimensions D and next restrict ourselves to three dimensions.

The generators of Poincar\'{e} group are translations$\mathcal{P}_{\mu }$, Lorentz rotations$\mathcal{J^{\mu \nu }}$ such that
\begin{eqnarray}\begin{split}
& [\mathcal{P}^{\mu }, \mathcal{P}^{\nu }]=0, ~~ [\mathcal{P}^{\mu }, \mathcal{J}^{\rho \sigma }]=i\left( \eta ^{\mu \sigma }\mathcal{P}^{\rho } -\eta ^{\mu \rho }\mathcal{P}^{\sigma } \right) , \\
& [\mathcal{J} ^{\mu \nu } , \mathcal{J} ^{\rho \sigma } ] = i\left( \eta ^{\nu \sigma }\mathcal{J}^{\mu \rho } -\eta ^{\nu \rho }\mathcal{J}^{\mu \sigma } -\eta ^{\mu \sigma }\mathcal{J}^{\nu \rho } +\eta ^{\mu \rho }\mathcal{J}^{\nu \sigma } \right) .
\end{split}\label{eq:a2-1}
\end{eqnarray}
In light-cone base, minding $\eta _{+-}=\eta ^{+-}=1$ and $\eta _{IJ}=\delta _{IJ}$, we find the following commutation relations:
\begin{eqnarray}\begin{split}
& [\mathcal{P}^{\pm }, \mathcal{J}^{+-}] =\pm i\mathcal{P}^{\pm }, ~~ [\mathcal{P}^{\pm }, \mathcal{J}^{\mp I}]=-i\mathcal{P}_{I}, \\
& [\mathcal{P}^{I}, \mathcal{J}^{\pm J}]=i\delta ^{IJ}\mathcal{P}^{\pm }, ~~ [\mathcal{P}^{I}, \mathcal{J}^{JK}]=i\left( \delta ^{IK}\mathcal{P}^{J} -\delta ^{IJ}\mathcal{P}^{K} \right) , \\
& [\mathcal{J} ^{+-} , \mathcal{J} ^{\pm I} ] =\mp i\mathcal{J}^{\pm I}, ~~ [\mathcal{J} ^{+I} , \mathcal{J} ^{-K} ] =i\left( \mathcal{J}^{IK}+\delta ^{IK}\mathcal{J}^{+-} \right) , \\
& [\mathcal{J} ^{\pm I} ,\mathcal{J} ^{KL} ] =i\left( \delta ^{IL}\mathcal{J}^{\pm K}-\delta ^{IK}\mathcal{J}^{\pm L} \right) , \\
& [\mathcal{J} ^{IJ} , \mathcal{J} ^{KL} ] = i\left( \delta ^{JL}\mathcal{J}^{IK} -\delta ^{JK}\mathcal{J}^{IL} -\delta ^{IL}\mathcal{J}^{JK} +\delta ^{IK}\mathcal{J}^{JL} \right) , \\
& \mbox{with others vanishing}, 
\end{split}\label{eq:a2-2}
\end{eqnarray}
where $I,J,K,L=2,\cdots ,D-1$.
Among commutators which must be zero, particularly
\begin{eqnarray}
[\mathcal{J}^{-I},\mathcal{J}^{-J}]  \overset{?}{=} 0 
\label{eq:a2-3}
\end{eqnarray}
cannot be satisfied when we quantize the string theory except for in three dimensions. This is so called a "dangerous commutator". 

\subsubsection*{3D}
In three dimensions Lorentz generators are rewritten as $\mathcal{J}^{\pm }= \mp \mathcal{J}^{\pm 2}, ~ \mathcal{J} =-\mathcal{J}^{+-}$ like eq. (\ref{eq:a1-5}). Hence the commutation relations (\ref{eq:a2-1}) become simple as follows, 
\begin{eqnarray}
[\mathcal{P}^{\mu }, \mathcal{P}^{\nu }]=0, ~~ [\mathcal{J}^{\mu }, \mathcal{P}^{\nu }]=i\epsilon ^{\mu \nu \rho }\mathcal{P}_{\rho } , 
~~ [\mathcal{J} ^{\mu } , \mathcal{J} ^{\nu } ] = i\epsilon ^{\mu \nu \rho }\mathcal{J}_{\rho } .
\label{eq:a2-4}
\end{eqnarray}
Moreover we can make two Poincar\'{e} Casimir operators easily as follow,
\begin{eqnarray}
M^2=-\mathcal{P}^2, ~~~ \Lambda =\mathcal{P}_{\mu }\mathcal{J}^{\mu }
\label{eq:a2-5}
\end{eqnarray}
Unitary irreducible representations of Poincar\'{e} group are labeled by the value of these two Casimirs\cite{bib:008} and particularly irreps. with $M^2\geq 0$ are only physical. When $M^2>0$ we define relativistic helicity by
\begin{eqnarray}
s=\frac{\Lambda }{M}
\label{eq:a2-6}
\end{eqnarray}
This may take either sign, and parity flips the sign of $s$. Further we call $|s|$ "spin". 
If Lorentz group is $SO(1,2)$, its double cover $SL(2;\mathbb{R})$ or its universal cover, $s$ is an integer, half-integer or any real number.

In light-cone base commutation relations of 3d Poincar\'{e} group are
\begin{eqnarray}\begin{split}
& [\mathcal{J}^{\pm }, \mathcal{P}^{\mp }]=\pm i \mathcal{P} , ~~ [\mathcal{J}, \mathcal{P}^{\pm }]=\pm i \mathcal{P}^{\pm }, ~~[\mathcal{J}^{\pm }, \mathcal{P}]=\mp i \mathcal{P}^{\pm } , \\
& [\mathcal{J} , \mathcal{J} ^{\pm } ] = \pm i\mathcal{J}^{\pm } , ~~  [\mathcal{J}^{+} , \mathcal{J} ^{-} ] = i\mathcal{J} , ~~~\mbox{with others vanishing}.
\end{split}\label{eq:a2-7}
\end{eqnarray}
In three dimensions the commutator $[\mathcal{J}^{-}, \mathcal{J}^{-}]$, corresponding to a dangerous commutator ($\ref{eq:a2-3}$), vanish trivially because the transverse direction is only one. 
Therefore the 3d string theory in Light-cone gauge has no Lorentz anomaly and is thought of preserving the Poincar\'{e} symmetry.

\subsection{Conformal group}
We expect a tensionless string to have the space-time conformal symmetry.
First we consider in general dimensions D and next in three dimensions.

The generators of the space-time conformal group are dilatation $\mathcal{D}$ and special conformal transformation $\mathcal{K}^{\mu }$, additional to translations$\mathcal{P}_{\mu }$ and Lorentz rotations$\mathcal{J^{\mu \nu }}$, such that
\begin{eqnarray}\begin{split}
& [\mathcal{D},\mathcal{P}^{\mu }]=i\mathcal{P}^{\mu } , ~~ [\mathcal{D},\mathcal{J}^{\mu \nu }]=0, ~~  [\mathcal{D},\mathcal{K}^{\mu }]=-i\mathcal{K}^{\mu } , ~~ [\mathcal{K}^{\mu }, \mathcal{K}^{\nu }]=0, \\
& [\mathcal{K}^{\mu } , \mathcal{P} ^{\nu } ] =i\left( \eta ^{\mu \nu }\mathcal{D}+\mathcal{J}^{\mu \nu } \right) , ~~ [\mathcal{K}^{\mu } , \mathcal{J} ^{\rho \sigma } ] = i\left( \eta ^{\mu \sigma }\mathcal{K}^{\rho } -\eta ^{\mu \rho }\mathcal{K}^{\sigma } \right) .
\end{split}\label{eq:a2-8}
\end{eqnarray}
In light-cone base, 
\begin{eqnarray}\begin{split}
& [\mathcal{D},\mathcal{P}^{\pm }]=i\mathcal{P}^{\pm } , ~~ [\mathcal{D},\mathcal{P}^{I}]=i\mathcal{P}^{I} , ~~  [\mathcal{D},\mathcal{K}^{\pm }]=-i\mathcal{K}^{\pm } , ~~  [\mathcal{D},\mathcal{K}^{I}]=-i\mathcal{K}^{I} , \\
& [\mathcal{K}^{\pm } , \mathcal{P}^{\mp } ] =i\left( \mathcal{D} \pm \mathcal{J}^{+-} \right) , ~~ [\mathcal{K}^{\pm } , \mathcal{P} ^{I} ] = -[\mathcal{K}^{I} , \mathcal{P} ^{\pm } ] =i\mathcal{J}^{\pm I} , \\
& [\mathcal{K}^{I} , \mathcal{P} ^{J} ] =i \left( \delta ^{IJ} D+\mathcal{J}^{IJ} \right) , \\
& [\mathcal{K}^{\pm } , \mathcal{J} ^{+-} ] = \pm i \mathcal{K}^{\pm }, ~~ [\mathcal{K}^{\pm }, \mathcal{J}^{\mp I}]=-i\mathcal{K}^{I}, ~~ [\mathcal{K}^{I}, \mathcal{J}^{\pm J}]=-i\delta ^{IJ}\mathcal{K}^{\pm }, \\
& [\mathcal{K}^{I}, \mathcal{J}^{JK}]=i\left( \delta ^{IL}\mathcal{K}^{K}-\delta ^{IK}\mathcal{K}^{L} \right) , ~~ \mbox{with others vanishing}. 
\end{split}\label{eq:a2-9}
\end{eqnarray}
A tensionless string in generic dimensions can have a dangerous commutator as well as the case of Lorentz anomaly. 
That is as follows:
\begin{eqnarray}
[\mathcal{K}^{I}, \mathcal{J}^{-J}] \overset{?}{=} -i \delta ^{IJ}\mathcal{K}^{-}.
\label{eq:a2-10}
\end{eqnarray}
The right hand side of this commutator has the off-diagonal, traceless part with respect to $I,J$ as well as the trace part. 
The difference of trace part can be absorbed in the redefinition of $\mathcal{K}^{-}$. 
However the traceless part remain as anomaly \cite{bib:007}.

\subsubsection*{3D}
In three dimensions the commutation relations (\ref{eq:a2-8}) become simple a little as follows,
\begin{eqnarray}\begin{split}
& [\mathcal{D},\mathcal{P}^{\mu }]=i\mathcal{P}^{\mu } , ~~ [\mathcal{D},\mathcal{J}^{\mu }]=0, ~~  [\mathcal{D},\mathcal{K}^{\mu }]=-i\mathcal{K}^{\mu } , ~~ [\mathcal{K}^{\mu }, \mathcal{K}^{\nu }]=0, \\
& [\mathcal{K}^{\mu } , \mathcal{P} ^{\nu } ] =i\left( \eta ^{\mu \nu }D-\epsilon ^{\mu \nu \rho }\mathcal{J}_{\rho } \right) , ~~ [\mathcal{K}^{\mu } , \mathcal{J} ^{\nu } ] = i\epsilon ^{\mu \nu \rho } \mathcal{K}_{\rho } .
\end{split}\label{eq:a2-11}
\end{eqnarray}
In light-cone base, 
\begin{eqnarray}\begin{split}
& [\mathcal{D},\mathcal{P}^{\pm }]=i\mathcal{P}^{\pm } , ~~ [\mathcal{D},\mathcal{P}]=i\mathcal{P} , ~~ [\mathcal{D},\mathcal{K}^{\pm }]=-i\mathcal{K}^{\pm } , ~~ [\mathcal{D},\mathcal{K}]=-i\mathcal{K} \\
& [\mathcal{K}^{\pm } , \mathcal{P}^{\mp } ] =i\left( \mathcal{D} \mp \mathcal{J} \right) , ~~ [\mathcal{K}^{\pm } , \mathcal{P} ] =-[\mathcal{K} , \mathcal{P}^{\pm } ] =\pm i\mathcal{J}^{\pm } , ~~ [\mathcal{K} , \mathcal{P} ] =i\mathcal{D} , \\
& [\mathcal{K}^{\pm } , \mathcal{J} ^{\mp } ] = \pm i \mathcal{K} , ~~ [\mathcal{K}^{\pm } , \mathcal{J} ] = \mp i \mathcal{K}^{\pm } , ~~ [\mathcal{K} , \mathcal{J} ^{\pm } ] = \pm i \mathcal{K}^{\pm } , \\
& \mbox{with others vanishing}.
\end{split}\label{eq:a2-12}
\end{eqnarray}
In three dimensions the commutator $[\mathcal{K}, \mathcal{J}^{-}]$, corresponding to a dangerous commutator ($\ref{eq:a2-10}$), is only one commutator and is regarded as the redefinition of $\mathcal{K}^{-}$. 
Therefore this type of anomaly does not exist. However there are many commutators not to be thought in tensile string theory and we must check that they satisfy the expected commutation relations of the conformal group. 
The calculation of them is very complicated and lengthy, in particular the next commutation relation is very troublesome;  
\begin{eqnarray}
[\mathcal{K}^{-},\mathcal{J}^{-}]=0.
\label{eq:a2-13}
\end{eqnarray}
A check of this relation is the main result of this paper.

\section{Normalization of the wave functions}
In section 4 we have investigated eigenfunctions of mass squared operator $\mathcal{M}^2$. In this section we give the detail with respect to the normalization of the wave functions \cite{bib:016}. 
The inner product of two total eigenfunctions (\ref{eq:4-16}) are defined with the following definition for each $n$. 

First we check eq.(\ref{eq:4-13}).
By the definition (\ref{eq:4-12}), the scalar product of $\psi _{m,s}$ and $\psi _{m',s'}$ is
\begin{eqnarray}
(\psi _{m,s },\psi _{m',s'}) = 2\pi \delta _{s,s'} N_{m}^{*} N_{m'} \int _{0}^{\infty } dr r J_{s}(2mr) J_{s}(2m'r) .
\label{eq:a3-1}
\end{eqnarray}
The integral in (\ref{eq:a3-1}) can be computed by using the following result
\begin{eqnarray}
\int _{0}^{y} dx x J_{l}(ax)J_{l}(bx) = \frac{y}{a^2 -b^2} \left[ a J_{l+1}(ay) J_{l}(by) -b J_{l}(ay) J_{l+1}(by) \right] ,
\label{eq:a3-2}
\end{eqnarray}
where $a$ and $b$ are positive.
We take a limit $y=\Lambda \rightarrow \infty $ of eq.(\ref{eq:a3-2}) and obtain 
{\setlength\arraycolsep{1pt}\begin{eqnarray}
\int _{0}^{\infty } dx x J_{l}(ax)J_{l}(bx) &=& \lim _{\Lambda \rightarrow \infty } \frac{\Lambda }{a^2 -b^2} \left[ a J_{l+1}(a\Lambda ) J_{l}(b\Lambda ) -b J_{l}(a\Lambda ) J_{l+1}(b\Lambda ) \right] \nonumber \\
&=& \frac{1}{\pi } \frac{1}{\sqrt{a b}} \lim _{\Lambda \rightarrow \infty } \left[ \frac{\sin (a-b)\Lambda }{a-b} -(-1)^l \frac{\cos (a+b)\Lambda }{a+b} \right] \nonumber \\
&=& \frac{1}{a}\delta (a-b) \ \ \ \mbox{ for } a>0 \ \mbox{and } b>0,
\label{eq:a3-3}
\end{eqnarray}}
where we used $J_{l}(0)=0$ for $l>0$ and the Hankel asymptotic form
\begin{eqnarray}
J_{\nu }(x) = \sqrt{\frac{2}{\pi x}} \left[ \cos \left( x-\frac{2\nu +1}{4}\pi \right) +\mathcal{O}(x^{-1}) \right] \ \ \mbox{as } x\rightarrow \infty 
\label{eq:a3-4}
\end{eqnarray}
and the delta function defined by the weak limit\footnote{Note that the delta function defined in this way makes sense only for the smooth function with compact support. In other word, on the interval of integration with zero, $\int dx \lim _{\Lambda \rightarrow \infty } \frac{\sin (\Lambda x)}{\pi x} f(x) =\int dx \lim _{\Lambda \rightarrow \infty } \frac{e^{i\Lambda x}-e^{-i\Lambda x}}{2 \pi i x} f(x) =f(0)$ for any smooth functions $f(x)$ with compact support. In our case, we suppose that this delta function and the smooth function together will be integrated.}
\begin{eqnarray}
\lim _{\Lambda \rightarrow \infty } \frac{\sin (\Lambda x)}{\pi x} \equiv \delta (x) .
\label{eq:a3-5}
\end{eqnarray}
Thus we obtain 
\begin{eqnarray}
(\psi _{m,s },\psi _{m',s'}) = \frac{\pi }{2} \delta _{s,s'} \frac{|N_{m}|^2 }{m} \delta (m-m') .
\label{eq:a3-6}
\end{eqnarray}
If $|N_{m}|=\sqrt{\frac{2m}{\pi }}$, we get $(\psi _{m,s },\psi _{m',s'}) = \delta _{s,s'} \delta (m-m')$.

Next we check the orthogonality between massive eigenfunctions and massless eigenfunctions.
For massless $m=0$, we consider the $s\not= 0$ case and the $s=0$ case separately. 

The general solution of eq.(\ref{eq:4-14}) for $s\not= 0$ is 
\begin{eqnarray}
\psi _{0,s}(r,\theta ) = (A r^{-|s|} +B r^{|s|}) e^{i s \theta }, 
\label{eq:a3-7}
\end{eqnarray}
where $A$ and $B$ are constants. 
For simplicity, we consider the $s>0$ case. The $s<0$ case is also discussed similarly.  
The scalar product of $\psi _{0,s}$ and $\psi _{m,s'}$ for $s>0$ is
{\setlength\arraycolsep{1pt}\begin{eqnarray}
(\psi _{0,s },\psi _{m,s'}) &=& 2\pi \delta _{s,s'} N_{m} \int _{0}^{\infty } dr \ r \ (A^{*} r^{-s} +B^{*} r^{s} ) J_{s}(2mr) \nonumber \\
&=& 2\pi \delta _{s,s'} N_{m} \left[ A^{*} \frac{m^{s-2}}{2 (s-1)!} +B^{*} \sqrt{\frac{2}{\pi }} \lim _{\Lambda \rightarrow \infty } \frac{\Lambda ^{s+\frac{1}{2}}}{(2m)^{\frac{3}{2}}} \cos \left( 2m\Lambda -\frac{2s+3}{4} \pi \right) \right] ,
\label{eq:a3-8}
\end{eqnarray}}
where we use the asymptotic form (\ref{eq:a3-4}) and the following results for positive integer $l$
{\setlength\arraycolsep{1pt}\begin{eqnarray}
\begin{split}
\int _{0}^{y} dx x^{l+1} J_{l}(x) &= y^{l+1} J_{l+1}(y) \\
\int _{0}^{y} dx x^{-l+1} J_{l}(x) &= -y^{-l+1} J_{l-1}(y) +\frac{1}{2^{l-1} (l-1)!} .
\end{split}
\label{eq:a3-9}
\end{eqnarray}}
For $m \not= 0$, the second term of eq.(\ref{eq:a3-8}) is zero because of the same reason as the definition of the delta function using $sinc$ function. 
Thus the r.h.s. of eq.(\ref{eq:a3-8}) vanishes only if $A=0$. 
Therefore, the solution eq.(\ref{eq:4-14}) for $s>0$ is $\psi _{0,s} = r^{s} e^{i s \theta }$.
In the same way, we find that the solution for $s<0$ is $\psi _{0,s} = r^{-s} e^{i s \theta }$.

The general solution of eq.(\ref{eq:4-14}) for $s=0$ is
\begin{eqnarray}
\psi _{0,0}(r,\theta ) = A \log r +B ,
\label{eq:a3-10}
\end{eqnarray}
where $A$ and $B$ are constants.
The scalar product of $\psi _{0,0}$ and $\psi _{m,s}$ is
{\setlength\arraycolsep{1pt}\begin{eqnarray}
(\psi _{0,0},\psi _{m,s}) &=& 2\pi \delta _{s,0} N_{m} \int _{0}^{\infty } dr \ r \ (A^{*} \log r +B^{*} ) J_{s}(2mr) \nonumber \\
&=& 2\pi \delta _{s,s'} N_{m} \left[ -\frac{A^{*}}{(2m)^2 } +\lim _{\Lambda \rightarrow \infty }(A^{*}\log \Lambda +B^{*}) \sqrt{\frac{2}{\pi }} \frac{\Lambda ^{\frac{1}{2}}}{(2m)^{\frac{3}{2}}} \cos \left( 2m\Lambda -\frac{3}{4} \pi \right) \right] ,
\label{eq:a3-11}
\end{eqnarray}}
where we use the asymptotic form (\ref{eq:a3-4}) , eq.(\ref{eq:a3-9}) and the following result
\begin{eqnarray}
\int _{0}^{y} dx \ x \log x \ J_{0}(x) = y J_{1}(y) +J_{0}(y)-1 .
\label{eq:a3-12}
\end{eqnarray}
For $m \not= 0$, the second term of eq.(\ref{eq:a3-11}) is zero again and the r.h.s. of eq.(\ref{eq:a3-11}) vanishes only if $A=0$. 
Therefore the solution of eq.(\ref{eq:4-14}) for $s=0$ is a constant.

If we collect these three cases, we obtain
\begin{eqnarray}
\psi _{0,s}(r,\theta ) = B r^{|s|} e^{i s \theta }.
\label{eq:a3-100}
\end{eqnarray}

\end{document}